\documentclass[twocolumn,showpacs,preprintnumbers,amsmath,amssymb,superscriptaddress,nofootinbib,pra]{revtex4-1}

\usepackage{graphicx}
\usepackage{amsmath}
\usepackage{epsfig}
\usepackage{bm}
\usepackage{cases}
\usepackage{empheq}
\usepackage[usenames]{color}

\def\>{{\rangle}}
\def\<{{\langle}}

\begin{document}

\title{ Higher-order Time-Symmetry-Breaking Phase Transition due to meeting of  an Exceptional Point and Fano Resonance}

\author{Satoshi \surname{Tanaka}}
\email{stanaka@p.s.osakafu-u.ac.jp}
\affiliation{Department of Physical Science, Osaka Prefecture
University, Gakuen-cho 1-1, Sakai 599-8531, Japan}
\author{Savannah  \surname{Garmon}}
\affiliation{Department of Physical Science, Osaka Prefecture
University, Gakuen-cho 1-1, Sakai 599-8531, Japan}
\author{Kazuki \surname{Kanki}}
\affiliation{Department of Physical Science, Osaka Prefecture
University, Gakuen-cho 1-1, Sakai 599-8531, Japan}
\author{Tomio Petrosky}
\affiliation{Center for Studies in Statistical Mechanics and Complex
Systems, The University of Texas at Austin, Austin, TX 78712 USA}
\affiliation{Institute of Industrial Science, The University of Tokyo, Tokyo 153-8505, Japan}

\begin{abstract}
We have theoretically investigated the time-symmetry breaking phase transition process for two discrete states coupled with a one-dimensional continuum by solving the nonlinear eigenvalue problem for the effective Hamiltonian associated with the discrete spectrum.
We obtain the effective Hamiltonian with use of the Feshbach-Brillouin-Wigner projection method.
Strong energy dependence of the self-energy appearing in the effective Hamiltonian plays a key role in the time-symmetry breaking phase transition: as a result of  competition in the decay process between the Van Hove singularity and the Fano resonance, the phase transition becomes a higher-order transition when both the two discrete states are located near the continuum threshold.

\end{abstract}
\date{\today}

\maketitle

\section{Introduction}\label{Sec:Intro}

In open systems, the time symmetry of the evolution equations such as the Schr{\"o}dinger equation is spontaneously broken with the appearance of  Poincar{\'e}  resonance, and irreversible process emerges \cite{PrigogineEndCertain}.
 Time-symmetry breaking is ubiquitous in nature, {\it e.g.},  nuclear decay and spontaneous emission \cite{Messiah,CohenTannoudji}, but at a glance it seems inconsistent with the principle of microscopic dynamics, which dictates  the time evolution of a physical system to be time-reversible subject to unitary time evolution.
Therefore, since the birth of quantum mechanics, finding a consistent interpretation for irreversible phenomena in a unified theoretical framework has been a fundamental difficulty \cite{Dirac27,Weisskopf30,Heitler36,Friedrichs48,Sudarshan78,BohmBook,Petrosky91}.

Recently, various extensions of quantum mechanics written in terms of  a phenomenological non-Hermitian Hamiltonian have been proposed\cite{Hatano96,Hatano97,Bender98,Bender03,Heiss04,Rotter09,Rotter15}; for example,   it has been revealed that if we weaken the standard requirement of Hermiticity in favor of  Parity-Time-symmetry (PT-symmetry) \cite{Bender98,Bender03} the effective Hamiltonian may exhibit a {\it bifurcation} of the eigenvalues from real to complex, as the system control parameter surpasses  a critical value.
This singular critical point is mathematically identified as an exceptional point (EP) where the effective Hamiltonian takes a Jordan block structure as not only the eigenvalues but also the eigenstates coalesce\cite{KatoBook,Heiss98,Heiss04,Rotter09}.

 This process can be viewed as a phase transition in a dynamical sense: below the critical point,  time-evolution is reversible, while irreversibility appears beyond the critical point.
Therefore we call this  {\it a Time-Symmetry Breaking  Phase Transition} (TSBPT) where the imaginary part of the eigenvalue can be regarded as an order parameter that has a singularity at the critical point in terms of a system control parameter\cite{Heiss98,Jung99,Pastawsky07,Rotter10,Bender13,Eleuch15,Garmon12,Brody13}.
Very recently TSBPT has been experimentally observed in mesoscopic quantum systems\cite{Alvarez06,Rotter15}, and also in many analogous optical systems where interesting collective dynamical properties have been  studied, such as superradiance and lasing \cite{Guo09,Ruter10,Peng14,Feng14}.
  
In the description of open quantum systems, a non-Hermitian effective Hamiltonian can be derived from the  microscopic  total (Hermitian) Hamiltonian including the environment with use of the Feshbach-Projection-Operator method (FPO method) without relying upon phenomenological equations \cite{Feshbach,Jung99,Rotter09,Rotter15,Rotter10,Eleuch15}, where  detailed information about the  microscopic interaction with the environment is renormalized into the self-energy, which is represented by a Cauchy integral in which the direction of the analytic continuation across the branch cut determines the direction of the arrow of time.
Prigogine and one of the authors (T.P.) {\it et al.} have clarified that the spectrum of the effective Hamiltonian coincides with that of the total Hamiltonian, so that the Hermitian Hamiltonian of the total system can have a complex spectrum due to the resonance if we extend the eigenvector space from the ordinary Hilbert space into a dual vector space, called the extended Hilbert space, where the Hilbert norm of the eigenvector vanishes\cite{BohmBook,Petrosky91,Petrosky97}.

An important feature of the effective Hamiltonian thus derived is that it may have  strong energy dependence, especially around the branch point.
As a result, the eigenvalue problem of the effective Hamiltonian is nonlinear in the sense that the operator itself depends on the energy eigenvalue.
This nonlinearity plays a crucial role in the TSBPT, because the self-energy  changes dramatically around the branch point bifurcation.
However, the role of this nonlinearity in the study of TSBPT has not been fully discovered yet.
As an example demonstrating the importance of the nonlinear effect, in our previous study on the decay process of an impurity in a one-dimensional (1D) conduction band we showed that the Van-Hove singularity in the density of states results in a strong non-analytic enhancement of the decay rate\cite{Tanaka06,GarmonPhD,Garmon09}.

In this paper, we consider a microscopic model consisting of  multiple discrete states coupled with a common 1D continuum, in which the nonlinearity plays an essential role in the properties of  the TSBPT.
As a result of the coupling of several resonance states through a common continuum, the individual decay processes interfere with each other yielding a characteristic spectral profile known as a Feshbach-Fano resonance\cite{Feshbach,Fano61,Joe06,Miroshnichenko10,Yoon12,Heiss14}.
As a result of the interference, the decay process may be completely suppressed for certain parameter values, which is also known as a bound state in continuum (BIC)\cite{Rotter09,Plotnik11,Weinmann13,Boretz14}.
Here we show that when the discrete state energies  are located near the branch point, the stabilization by the Fano resonance and the destabilization by the Van Hove singularity  compete to introduce entirely new dynamics.

Though several works have investigated the TSBPT associated with  multi-level systems coupling with a common continuum \cite{Heiss98,Jung99,Pastawsky07,Rotter09,Rotter10,Rotter15,Eleuch15}, the nonlinear effect has  been examined in less detail.
In this work, we reveal that the usual {\it second-order}  phase transition becomes  {\it fourth-order}  when two resonant states appear near the continuum threshold, by carefully considering the nonlinear effect of the microscopic effective Hamiltonian.
Furthermore, we find that the decaying state is more strongly stabilized due to  this competition than the stabilization resulting from an ordinary Fano resonance.

In Section~\ref{Sec:Model}, we present our model  and the nonlinear eigenvalue problem of the effective Hamiltonian by use of the FPO method.
Before studying the decay process for two discrete states, in Section~\ref{Sec:Single} the nonanalytical enhancement of the decay process due to the Van Hove singularity is briefly reviewed for a single state model. 
The main results of this paper for the two discrete states model are presented in Section~\ref{Sec:Double}, where it is revealed that the order of the TSBPT is significantly modified as a result of the competition between the Van Hove singularity and the Fano resonance.
Section~\ref{Sec:Discussion} is devoted to clarifying the role of  the Van Hove singularity by comparing these results to the decay process for a  three-dimensional (3D) system; we also propose some experiments to observe our findings.
 In Appendix~\ref{AppSec:Jordan}, we  heuristically present an effective non-Hermitian Hamiltonian which elucidates the system properties at the EP.

\section{Model and effective Hamiltonian} \label{Sec:Model}

\begin{figure}
\begin{center}
\includegraphics[height=5cm,width=4cm]{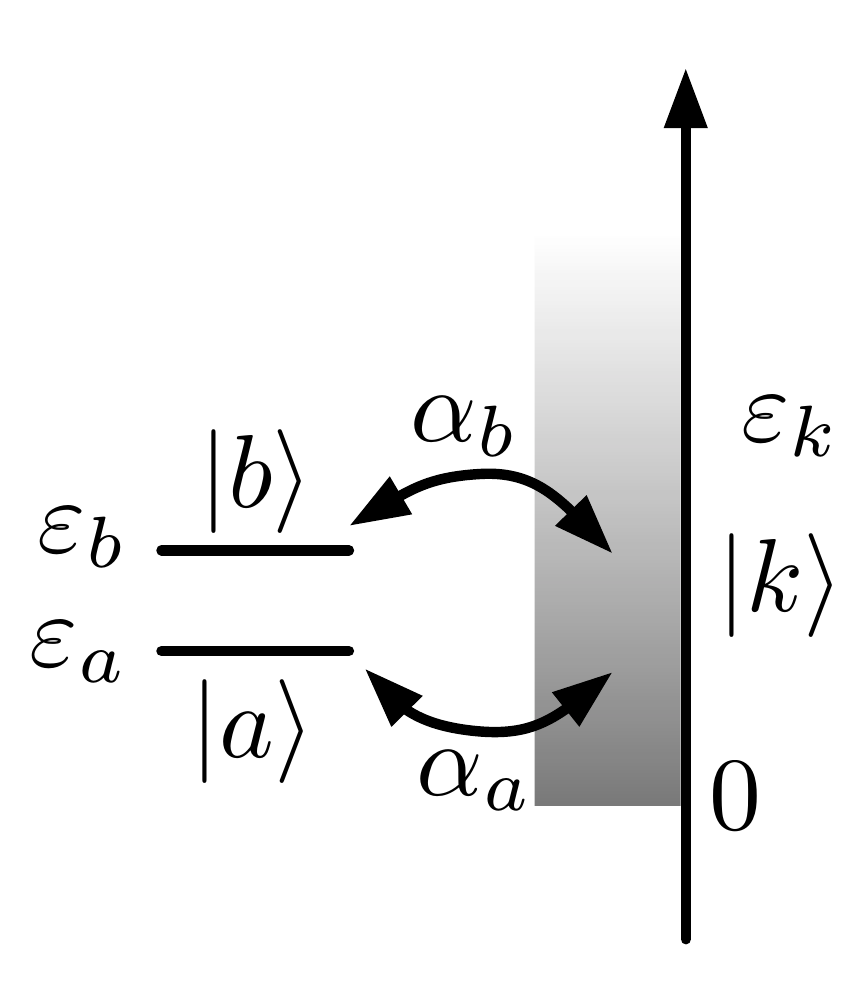}
\caption{Two discrete states $|a\>$ and $|b\>$ coupled with a 1D continuum $|k\>$. }
\label{Fig_model}
\end{center}
\end{figure}

We shall consider two discrete states $|a\>$ and $|b\>$ with their respective energies, $\varepsilon_a$ and $\varepsilon_b$, coupled with a one-dimensional continuous state $|k\>$ with the energy  $\varepsilon_k$ as shown in Fig.\ref{Fig_model}.
The Hamiltonian of the total system is given by
\begin{equation}\label{totalH}
\hat H=\hat H_0+\hat W  \;,
\end{equation}
where
\begin{subequations}\label{H0W}
\begin{eqnarray}
\hat H_0&=&\varepsilon_a |a\>\<a|+\varepsilon_b |b\>\<b|+\int_{-k_c}^{k_c} dk  \varepsilon_k |k\>\<k|  \;,\\
\hat W&=&\alpha_a\int_{-k_c}^{k_c} dk \left(|a\>\< k|+|k\>\<a|\right)\nonumber\\
&&+\alpha_b\int_{-k_c}^{k_c} dk \left(|b\>\<k|+|k\>\<b|\right) \;.
\end{eqnarray}
\end{subequations}
In Eq.(\ref{H0W}a), as a typical example of a continuous state with a Van Hove singularity at the continuum threshold, we consider the energy dispersion of a one-dimensional free particle represented by
\begin{equation}\label{ek}
\varepsilon_k ={\hbar^2 k^2 \over 2m} \;,
\end{equation}
while  $\alpha_a$ and $\alpha_b$ are the coupling strengths.
In this work, we choose units such that $\hbar=k_c=\omega_c=2m=1$, where $k_c$ and $\omega_c$ are the cut-off wavelength and frequency of the continuum.
With use of these units, all the parameters are dimensionless in this paper.

When the discrete states are in resonance with the continuum, they generally decay into the continuum.
In order to explain the exponential decay process in terms of the microscopic dynamics, we consider the complex eigenvalue problem in the extended Hilbert space\cite{Petrosky91}
\begin{equation}\label{EVH}
\hat H|\Psi_j\>=z_j|\Psi_j\>\;, \; \<\tilde\Psi_j|\hat H=z_j\<\tilde\Psi_j| \;,
\end{equation}
where $|\Psi_j\>$ and $\<\tilde\Psi_j|$ are the right- and left-eigenstates of $\hat H$ with a common complex eigenvalue $z_j$.
In order to solve the complex eigenvalue problem Eq.(\ref{EVH}), we shall use the FPO method with the projection operators defined by
\begin{equation}\label{Projection_ab}
\hat P\equiv |a\>\<a| +|b\>\<b| \;, \; \hat Q=1-\hat P \;.
\end{equation}
Acting with $\hat P$ and $\hat Q$  on the first equation of Eq.(\ref{EVH}), we have
\begin{subequations}\label{ProjEV}
\begin{eqnarray}
&&\hat P H_0 \hat P|\Psi_j\>+\hat P \hat W \hat Q|\Psi_j\>=z_j \hat P|\Psi_j\>   \;,\\
&&\hat Q \hat W\hat P|\Psi_j\>+\hat Q H \hat Q|\Psi_j\>=z_j \hat Q|\Psi_j\>   \;.
\end{eqnarray}
\end{subequations}
From the second equation, we have
\begin{equation}\label{Qcomp}
\hat Q|\Psi_j\>=\hat Q {1 \over z_j- \hat{Q}H \hat{Q} }  \hat Q\hat W \hat P |\Psi_j\>  \;.
\end{equation}
Substituting Eq.(\ref{Qcomp}) into Eq.(\ref{ProjEV}a) results in
\begin{equation}\label{HeffEV}
\hat H_{\rm{eff}}(z_j)\hat P|\Psi_j\>=z_j \hat P|\Psi_j\>  \;,
\end{equation}
where  $\hat H_{\rm{eff}}(z)$ is an effective Hamiltonian defined by
\begin{subequations}\label{Heff}
\begin{eqnarray}
\hat H_{\rm{eff}}(z) &\equiv&
 \hat{P} H_0  \hat{ P}
+\hat P\hat W\hat{ Q}{1 \over z- \hat{Q}H \hat{ Q}}  \hat{ Q}\hat W\hat{ P} \\
&=&\hat H_0+\hat \Sigma(z)  \;,
\end{eqnarray}
\end{subequations}
and $\hat\Sigma(z)$ is the energy dependent self-energy operator.
It should be emphasized that the spectrum of the effective Hamiltonian coincides with the discrete spectrum of the total Hamiltonian.

There are two important characteristics of the eigenvalue problem of $\hat H_{\rm{eff}}(z_j) $ in Eq.(\ref{HeffEV}).
First, the self-energy propagator possesses a resonance singularity, which renders $\hat H_{\rm{eff}}(z_j) $ non-Hermitian.
Second, the eigenvalue problem in Eq.(\ref{HeffEV}) is  nonlinear in the sense that  $\hat H_{\rm{eff}}(z_j) $ itself depends on the eigenvalue through the self-energy operator $\hat\Sigma(z)$, with which the eigenvalue must be self-consistently determined.

In the present case, the effective Hamiltonian is represented  in terms of  the $\{ |a\>,|b\>\}$-basis by
\begin{equation}\label{Heff_ab}
\hat H_{\rm{eff}}(z)
=\begin{pmatrix}
 \varepsilon_a  &0\\
0&\varepsilon_b  
\end{pmatrix}   
+\sigma(z)
\begin{pmatrix}
\alpha^2_a &\alpha_a \alpha_b \\
\alpha_a\alpha_b &\alpha_b^2
\end{pmatrix} \;,
\end{equation}
where  $\sigma(z)$ is the scalar self-energy defined by  
\begin{equation}\label{selfDef}
\sigma (z)\equiv {1\over 2 k_c}\int_{-k_c}^{k_c} {dk\over z-\varepsilon_k}   \;,
\end{equation}
with the cut-off wavenumber $k_c$ ($k_c=1$, as denoted above)  to avoid an ultraviolet divergence of the integral. 
With use of  Eq.(\ref{ek}), $\sigma(z)$ is given by
\begin{equation}\label{selfEng}
\sigma(z) ={1\over 4 k_c}\int_0^{k_c^2} {1 \over \sqrt{\varepsilon}}{d\varepsilon\over z-\varepsilon}=1-i{\pi \over 2\sqrt z}\;.
\end{equation}
The scalar self-energy $\sigma(z)$ expressed as a Cauchy integral has a branch cut along the positive real axis of $\varepsilon_k$ so that it becomes a two-valued complex function.
By analytic continuation $\sigma(z)$ becomes an analytic function in a two-sheet Riemann surface.
As seen in Eq.(\ref{Heff_ab}), the two discrete states are indirectly coupled to  each other via interactions with a common continuum.
The imaginary part of this off-diagonal element is essential to the Fano resonance as will be shown in Section \ref{Sec:Double}.
It should be noted that the self-energy is divergent  at the branch point $z=0$, which is due to the Van Hove singularity.
The Van Hove singularity introduces a number of non-analytic effects into the system, including an enhancement of the decay rate near the branch point\cite{Tanaka06,GarmonPhD,Garmon09,Petrosky05}.

The eigenvalue $z_j$ is obtained as a solution of the dispersion equation $\det(\hat H_{\rm eff}(z)-z\hat {I}) = 0$ from Eq.(\ref{HeffEV}), which is explicitly written as
\begin{eqnarray}\label{disp}
f(z;\varepsilon_a,\varepsilon_b)& \equiv&\left(z-\varepsilon_a-\alpha_a^2\sigma(z)\right)\left(z-\varepsilon_b-\alpha_b^2\sigma(z)\right)\nonumber\\
&&-\alpha_a^2\alpha_b^2\sigma^2(z)=0 \;.
\end{eqnarray}
When we rewrite this as, for example, 
\begin{equation}\label{disp2}
z=\varepsilon_a+\alpha_a^2\sigma(z) + { \alpha_a^2\alpha_b^2\sigma^2(z) \over  z-\varepsilon_b-\alpha_b^2\sigma(z)} \;,
\end{equation}
the physical meaning of each term of the {\it r.h.s.} is clear: the first term is the unperturbed energy of  the $|a\>$ state, the second term represents the direct interaction of the $|a\>$ state with the continuum, while the third term  represents the indirect coupling with the $|b\>$ state through the  continuum.

Substituting Eq.(\ref{selfEng}) into Eq.(\ref{disp}), the dispersion equation takes the form of a fifth-order polynomial equation
\begin{eqnarray}\label{Disp12}
&&f(z;\varepsilon_a,\varepsilon_b)= 4 z \left\{ (z-\varepsilon_a- \alpha_a^2) (z-\varepsilon_b- \alpha_b^2)- \alpha_a^2\alpha_b^2 \right\}^2\nonumber\\
&&\qquad+\pi^2 \left\{ (\alpha_a^2+\alpha_b^2)z-(\alpha_b^2\varepsilon_a+\alpha_a^2\varepsilon_b ) \right\}^2=0\;.
\end{eqnarray}
Since the scalar self-energy $\sigma(z)$ is analytically continued to the second Riemann sheet, the five solutions of Eq.(\ref{Disp12}) are located in either the first or  second Riemann sheet.
In Section~\ref{Sec:Double}, we see how the time-symmetry breaking transition occurs in the second sheet as the  parameters  $\varepsilon_a$ and $\varepsilon_b$ are varied.

There is another way to represent the effective Hamiltonian that is useful to describe the situation when $\varepsilon_a\simeq\varepsilon_b$.
Using the basis transformation 
\begin{subequations}\label{baseFano}
\begin{eqnarray}
&&|\psi_F\>={|a\>-|b\> \over\sqrt 2}  \;, \\
&&|\psi_{AF}\>={|a\>+|b\> \over \sqrt 2},
\end{eqnarray}
\end{subequations}
the effective Hamiltonian is represented by
\begin{equation}\label{Heff_FAF}
\hat H_{\rm{eff}}(z)
=\begin{pmatrix}
 \varepsilon_A  & \varepsilon_D\\
\varepsilon_D &\varepsilon_A 
\end{pmatrix}   
+\sigma(z) 
\begin{pmatrix}
\alpha^2_D &\alpha_A \alpha_D \\
\alpha_A\alpha_D &\alpha_A^2
\end{pmatrix} \;,
\end{equation}
where the average and the difference of the discrete state energies are respectively defined as
\begin{equation}\label{eneAD}
\varepsilon_A={\varepsilon_a+\varepsilon_b\over 2} \;,\; \varepsilon_D={\varepsilon_a-\varepsilon_b\over 2} \;,
\end{equation}
and the average and the difference of the coupling strengths are also respectively defined as 
\begin{equation}\label{alphaAS}
\alpha_A={\alpha_a+\alpha_b\over \sqrt{2}} \;,\; \alpha_D={\alpha_a-\alpha_b\over \sqrt{2}} \;.
\end{equation}
The dispersion equation in this representation reads as
\begin{eqnarray}\label{Disp_AS}
&&f(z;\varepsilon_A,\varepsilon_D)\equiv 4 z \bigg\{ (z-\varepsilon_A) \left(z-\varepsilon_A- (\alpha_A^2+\alpha_D^2) \right) \nonumber\\
&&\hspace{3cm} -\varepsilon_D(\varepsilon_D+2 \alpha_A\alpha_D)  \bigg\}^2 \nonumber\\
&&\quad  + \pi^2\bigg\{ (\alpha_A^2+\alpha_D^2 )  (z-\varepsilon_A ) + 2\alpha_A \alpha_D \varepsilon_D \bigg\}^2=0\;.
\end{eqnarray}
We find  that  $z=\varepsilon_A$ is a double real root of Eq.(\ref{Disp_AS}) when $\varepsilon_D=0$, i.e. $\varepsilon_a=\varepsilon_b$.
This indicates that the decay rate vanishes as a consequence of the destructive interference of the two decay channels  $|a\>$ and $|b\>$, an example of  Fano resonance.

Before studying the TSBPT in the two discrete system, we  briefly review the TSBPT of a single discrete state in the next section.

\section{  TSBPT in  single discrete state model: role of  the Van Hove singularity}\label{Sec:Single}

In this section, we study a single discrete state coupled with the continuum to show that the time-symmetry breaking bifurcation is  a second-order phase transition resulting from the nonlinearity in the eigenvalue problem of the effective Hamiltonian.

We consider a single discrete state $|a\>$  coupled with a 1D continuum, where the total Hamiltonian is given by
\begin{eqnarray}\label{H1}
\hat H_1 &&=\varepsilon_a |a\>\<a|+\int_{-k_c}^{k_c} dk  \varepsilon_k |k\>\<k| \nonumber\\
&& + \alpha \int_{-k_c}^{k_c} dk \left(|a\>\< k|+|k\>\<a|\right) \;.
\end{eqnarray}
Note that we again take $k_c=1$ as mentioned in Section~\ref{Sec:Model}.
Using the FPO method with $\hat P=|a\>\<a|$, we  obtain the effective Hamiltonian as
\begin{equation}\label{SingleHeff}
\hat H_{\rm{eff},1}(z)=\varepsilon_a+\alpha^2\sigma(z) \;,
\end{equation}
where it should be noted that the effective Hamiltonian is a scalar operator in this case because the subsystem consists of  a single state  $\{|a\>\}$.

With use of  Eq.(\ref{selfEng}) for $\sigma(z)$, the dispersion equation reads 
\begin{equation}
z=\varepsilon_a+\alpha^2\left(1-i{\pi \over 2\sqrt z}\right) \;,
\end{equation}
equivalent to the third-order polynomial equation 
\begin{equation}\label{SingleDisp}
f_1(z;\varepsilon_a)\equiv z  (z-\varepsilon_a- \alpha^2)^2+{\pi^2\alpha^4\over 4}=0 \;.
\end{equation}
Because of the nonlinearity of the eigenvalue problem in the effective Hamiltonian, we have three eigenvalues even with a one-dimensional subsystem.

In order to evaluate the bifurcation point, we simultaneously solve Eq.(\ref{SingleDisp}) and its derivative
\begin{equation}\label{g1}
g_1(z;\varepsilon_a)\equiv {d\over dz}f_1(z;\varepsilon_a)=3z^2-4(\varepsilon_a+\alpha^2)z +(\varepsilon_a+\alpha^2)^2=0  \;.
\end{equation}
The condition for a common solution of Eqs.(\ref{SingleDisp}) and (\ref{g1}) requires that the determinant of the  Sylvester matrix, i.e., resultant, should satisfy \cite{Akritas93,GarmonPhD}
\begin{equation}\label{Resul1}
{\rm res}(f_1(z;\varepsilon_a),g_1(z;\varepsilon_a))=0 \;.
\end{equation}

The location of  the bifurcation point $\varepsilon_{c,1}$ in the parameter space of $\varepsilon_a$ is easily obtained as
\begin{equation}\label{SingleEc}
\varepsilon_a=\varepsilon_{c,1}\equiv-3\left({\pi \alpha^2\over 4}\right)^{2/3} -\alpha^2  <0  \;,
\end{equation}
and the common eigenvalue at the bifurcation point is given by
\begin{equation}\label{SingleZc}
z_{c,1}=-\left({\pi \alpha^2\over 4}\right)^{2/3}  \;.
\end{equation}
The two eigenvalues coalesce at $\varepsilon_a=\varepsilon_{c,1}$ before becoming a complex conjugate pair for  $\varepsilon_a >\varepsilon_{c,1}$.

\begin{figure}[t]
\begin{center}
\includegraphics[width=8.5cm]{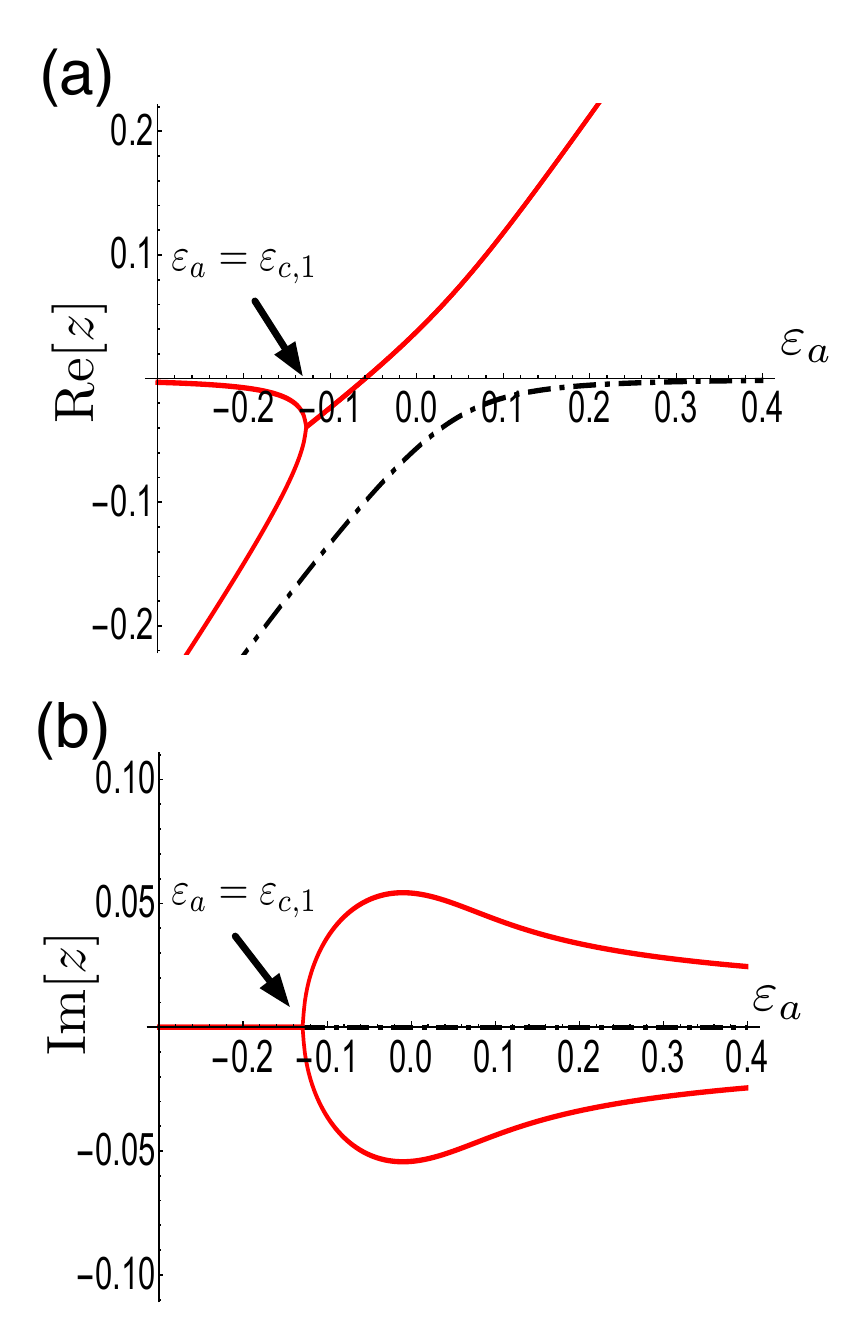}
\caption{(Color online) The eigenvalues of $\hat H_{\rm eff}$ as a function of $\varepsilon_a$ for the single discrete state system for $\alpha=0.1$. 
Real parts and imaginary parts are shown in the upper and lower panels, respectively.
The solid curves represent the solutions in the second Riemann sheet, while the dashed-dotted curve represents the PBS in the first Riemann sheet.
The arrow indicates the bifurcation point. }
\label{Fig:EVsingle}
\end{center}
\end{figure}

Next we examine the non-analytic properties of the spectrum near the bifurcation point, which are due to the influence of the nearby Van Hove singularity.
We show the eigenvalues as a function of $\varepsilon_a$  in Fig.\ref{Fig:EVsingle} for $\alpha=0.1$.
Since the dispersion equation is a third-order polynomial, there are three solutions of Eq.(\ref{SingleDisp}).
One real solution always exists in the first Riemann sheet below the band edge for any value of $\varepsilon_a$, which is a {\it persistent bound state} (PBS) attributed to the Van Hove singularity\cite{Tanaka06,Garmon09,Garmon13},  shown by the dashed-dotted curve in Fig.\ref{Fig:EVsingle}.
The other two solutions in the second Riemann sheet are bifurcated from two real solutions to a complex conjugate pair of  solutions at the bifurcation point, $\varepsilon_a=\varepsilon_{c,1}$, which are shown by the solid curves.

Now we  obtain an analytical expression of the solutions of $z$ in Eq.(\ref{SingleDisp}) around the bifurcation point.
Expanding $f_1(z;\varepsilon_a)$ around $z_{c,1}$  as a function of $p\equiv z-z_{c,1}$ and leaving  terms up to  order $p^2$, the dispersion equation (\ref{SingleDisp}) is written as
\begin{eqnarray}\label{Single3rdPol}
&&{1\over 4}\left( 3\alpha^{4/3} (2\pi)^{2/3} -8 u \right) p^2\nonumber\\
&&+{1\over 2}\left( -\alpha^{4/3}(2\pi)^{2/3} u+2 u^2 \right) p \nonumber\\
&&+ {1\over 2^{4/3}} \left(  \alpha^{8/3} (2\pi^2)^{2/3} u -\alpha^{4/3}\pi^{2/3} u^2 \right)=0  \;,
\end{eqnarray}
where $u\equiv \varepsilon_a-\varepsilon_{c,1}$ is the deviation from the bifurcation point in the parameter space.
Under the condition  
\begin{equation}\label{u_cond1}
u\ll O(\alpha^{4/3})  \;,
\end{equation}
 the solutions near the bifurcation point are approximately described by
\begin{equation}
p={u\over 3}\pm i \sqrt 3 (2\pi \alpha^2)^{1/3} \sqrt u \;.
\end{equation}
Therefore the two eigenvalues in this vicinity behave as
\begin{equation}\label{SingleZpm}
z_\pm(\varepsilon_a)={\varepsilon_a+\alpha^2 \over 3}\pm {i\over \sqrt 3}(2\pi\alpha^2)^{1/3}\sqrt{\varepsilon_a-\varepsilon_{c,1}} \;,
\end{equation}
where $\varepsilon_{c,1}$ is given by Eq.(\ref{SingleEc}).
The first-order derivative of $z_\pm(\varepsilon_a)$ in terms of $\varepsilon_a$ is discontinuous at the bifurcation point\cite{Garmon12,Garmon15} as shown in Fig.\ref{Fig:EVsingle}: in this sense, we can think of the time-symmetry breaking transition as a second-order phase transition.
Note also that the decay rate is proportional to $\alpha^{2/3}$ which is  non-analytically enhanced by the Van Hove singularity, compared to the ordinary decay rate determined by  Fermi's golden rule: $\alpha^{2/3} \gg \alpha^2$ for $|\alpha| <1$ \cite{Tanaka06}.

We have shown so far that  the two eigenvalues coalesce at the bifurcation point.
In Eq.(\ref{SingleZpm}), the eigenvalues are described by a fractional power expansion around $\varepsilon_a=\varepsilon_{c,1}$, indicating that this bifurcation point is an  EP, which is a singularity of a characteristic equation of a linear operator, at which the eigenstates  coalesce and the operator can be no longer diagonalized    \cite{KatoBook,KnoppBook}.
At the EP, the operator can only be reduced to Jordan block form. 
Even though our effective Hamiltonian is a scalar operator, we present in Appendix \ref{AppSec:Jordan}  an effective non-Hermitian Hamiltonian represented by two-by-two matrix which becomes a Jordan block matrix at the bifurcation point.
Therefore the bifurcation point at $\varepsilon_a=\varepsilon_{c,1}$ is consistent with an usual definition of an  EP \cite{EP2A}.

%

\section{TSBPT in Two discrete state model: Competition between the  EP and Fano resonance}
\label{Sec:Double}

In the preceding section, we have investigated  the TSBPT associated with a single discrete state in a 1D system, where it appears as a second order phase transition in which the transition region and the decay strength are exaggerated by Van Hove singularity.
In this section, we  shall clarify the effect of  the interaction between  resonance states on the TSBPT, by studying the two discrete system described in Section~\ref{Sec:Model}.
We especially focus on the competitive effects of the EP and the Fano resonance on  the TSBPT as mentioned in the Introduction.
In this section, we assume  $\alpha_a=\alpha_b=\alpha$ in Eq.(\ref{H0W}b), which does not change any essential physics as long as the interaction strengths are about  the same order of the magnitude.

\begin{figure}[!]
\begin{center}
\includegraphics[width=8.5cm]{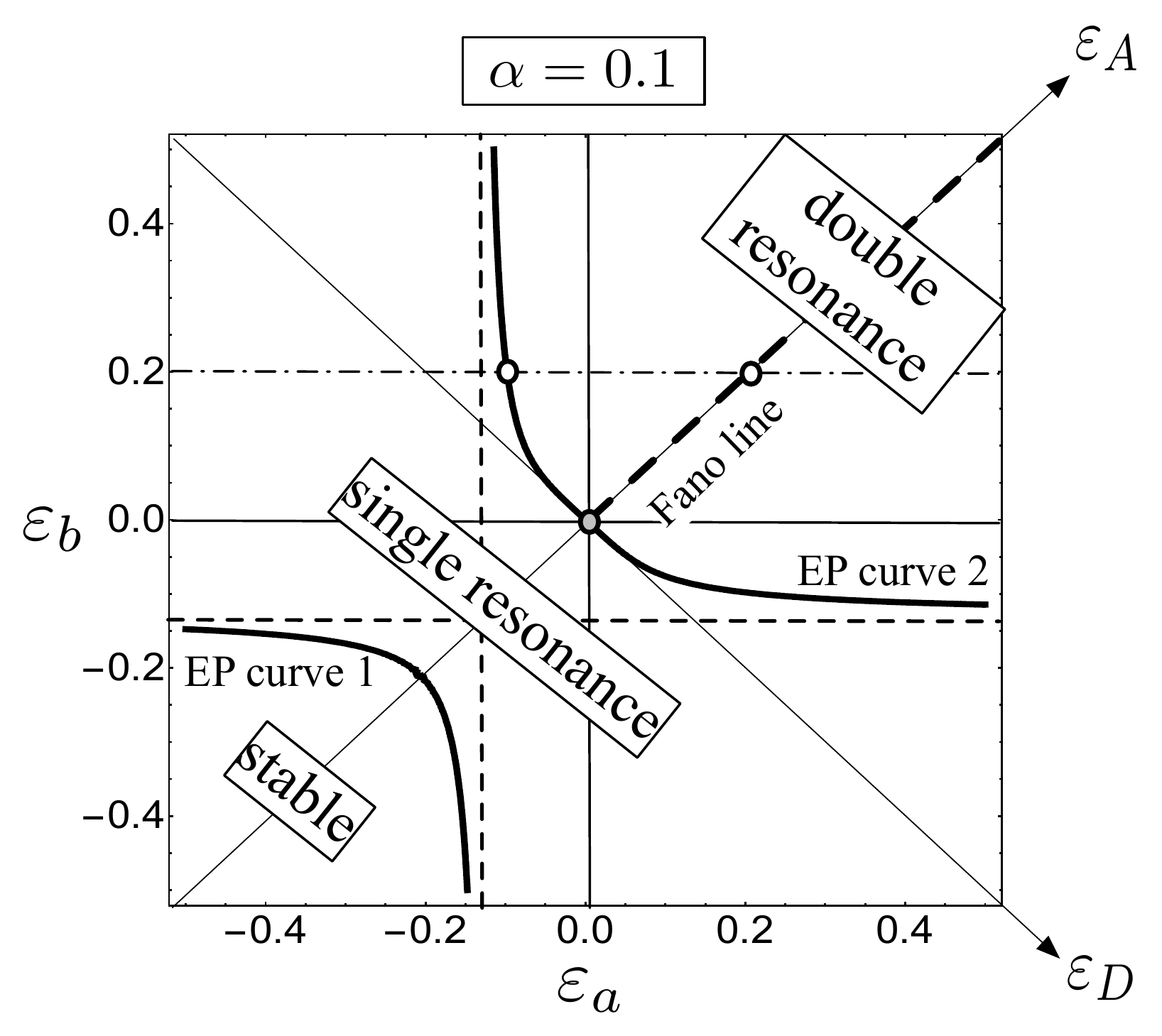}
\caption{ Phase diagram for the time-symmetry breaking transition for $\alpha=0.1$.
Solid lines represent the EP as a function of $\varepsilon_a$ and $\varepsilon_b$.
There are three phases: stable (five solutions are real eigenvalues), single resonance, and double resonance.
Each is separated by one of the two EP curves.
The thin dashed line represents the EP line in the single discrete state system: $\varepsilon_{a,b}=\varepsilon_{c,1}$.
The thick dashed line represents the Fano resonance.
The chain line at $\varepsilon_b=0.2$ corresponds to Fig.\ref{fig:EV}, and blank circles corresponds to  the EP and the Fano resonance.
The tilted axes in terms of $\varepsilon_A=(\varepsilon_a+\varepsilon_b)/2$ and $\varepsilon_D=(\varepsilon_a-\varepsilon_b)/2$ are shown by the thin solid lines.
The gray circle at the origin corresponds to the meeting point of the EP and the Fano resonance in Fig.\ref{fig:EPFano}. }
\label{fig:phase}
\end{center}
\end{figure}

Similar to the single state case in Eqs.(\ref{SingleDisp}) to (\ref{Resul1}), the exceptional point is obtained here as a common solution of Eq.(\ref{Disp12}) and its derivative
\begin{equation}
g(z;\varepsilon_a,\varepsilon_b)\equiv {d\over d z}f(z;\varepsilon_a,\varepsilon_b)=0  \;,
\end{equation}
which requires that the resultant is zero:
\begin{equation}\label{resulab}
{\rm res}\left(f(z;\varepsilon_a,\varepsilon_b),g(z;\varepsilon_a,\varepsilon_b)\right)=0  \;.
\end{equation}
This gives the exceptional point as a function of $\varepsilon_a$ and $\varepsilon_b$  as shown in Fig.\ref{fig:phase} for $\alpha=0.1$.
In Fig.\ref{fig:phase}, we denote the regions as  "{\it stable phase}" where all the solutions of  Eq.(\ref{Disp12}) are real,  "{\it single-resonance phase}" where we have a resonance and anti-resonance pair, and  "{\it double-resonance phase}" where we have two resonance and anti-resonance pairs. 

We now consider the condition for the  appearance of  the EP. 
The physical situation is different depending on the interaction strength between the two resonance states.
We first study the case where the effect of the interaction is weak and the two discrete energies $\varepsilon_a$ and $\varepsilon_b$ are far apart:
\begin{subequations}\label{cond1}
\begin{eqnarray}
&&|\varepsilon_b-\varepsilon_a|> \alpha \quad \text{and} \quad \varepsilon_b> 0 > \varepsilon_a \;, \\
\text{or   }\quad &&|\varepsilon_b-\varepsilon_a|> \alpha \quad \text{and}\quad  0>\varepsilon_b>\varepsilon_a  \;.
\end{eqnarray}
\end{subequations}

For the case of Eq.(\ref{cond1}a), dividing Eq.(\ref{Disp12}) by $\varepsilon_b^2$ yields (recall $\alpha_a=\alpha_b=\alpha$) 
\begin{eqnarray}\label{disp_eb}
&&z \left\{ (z-\varepsilon_a- \alpha^2) \left( {z-\alpha^2 \over \varepsilon_b} -1\right)- {\alpha^4\over \varepsilon_b} \right\}^2\nonumber\\
&&+{\pi^2\alpha^4\over 4} \left({2z-\varepsilon_a\over \varepsilon_b}-1 \right)^2=0\;.
\end{eqnarray}
Under the present case, we can neglect the term  $\alpha^4/\varepsilon_b$, which brings about
\begin{equation}\label{disp_eb2}
\tilde f(z;\varepsilon_a,\varepsilon_b) \equiv  z  (z-\varepsilon_a- \alpha^2)^2+{\pi^2\alpha^4\over 4} \xi^2(z,\varepsilon_a, \varepsilon_b )=0\;,
\end{equation}
where $\xi(z,\varepsilon_a,\varepsilon_b)$ is a correction from the interaction defined by
\begin{eqnarray}\label{xi}
\xi(z,\varepsilon_a,\varepsilon_b)&\equiv&{ 2z-\varepsilon_a - \varepsilon_b  \over  z-\alpha^2 - \varepsilon_b  }=1+{\varepsilon_a-z+\alpha^2 \over \varepsilon_b-z+\alpha^2} \nonumber\\
&\simeq&  1+{\varepsilon_a-z \over \varepsilon_b-z}  \;.
\end{eqnarray}
When the value of $z$ is close to the EP, we can replace $z$ in $\xi(z,\varepsilon_a,\varepsilon_b)$ with $z_{c1}$  to yield a similar   dispersion equation as the single discrete state system given by Eq.(\ref{SingleDisp}).
We find
\begin{equation}\label{disp_eb3}
\tilde f(z;\varepsilon_a,\varepsilon_b) \simeq   z  (z-\varepsilon_a- \alpha^2)^2+{\pi^2\tilde\alpha_1^4\over 4}=0\;,
\end{equation}
where $\tilde\alpha_1$ is the corrected interaction strength
\begin{equation}\label{correctionalpha}
\tilde\alpha_1^4\equiv \alpha^4 \left| 1+{\varepsilon_a-z_{c,1}\over \varepsilon_b-z_{c,1}}\right|^2   \;.
\end{equation}
Since for the case of Eq.(\ref{cond1}a) 
\begin{equation}
\tilde\alpha_1^4 < \alpha^4  \;,
\end{equation}
the effect of the interaction between the resonance states effectively reduces the interaction of the bare discrete state with the continuum so that the value of the EP  (EP curve 2) which divides the single-resonance  and the double-resonance phase lies closer to the continuum threshold than the value of the EP in the single discrete state case which is depicted by the dashed line at $\varepsilon_a=\varepsilon_{c,1}$ in Fig.\ref{fig:phase}:
\begin{equation}\label{EcApprox}
\varepsilon_{c,1}< \tilde\varepsilon_{c,1}\equiv-3\left({\pi \tilde\alpha_1^2\over 4}\right)^{2/3} -\tilde\alpha_1^2  <0  \;.
\end{equation}

On the other hand, for the case of  Eq.(\ref{cond1}b), by dividing Eq.(\ref{Disp12}) by $\varepsilon_a^2$, and repeating the above  procedure,  the dispersion equation is approximated as a third-order polynomial equation as
\begin{equation}\label{disp_eb4}
\tilde f(z;\varepsilon_a,\varepsilon_b) \simeq z  (z-\varepsilon_b- \alpha^2)^2+{\pi^2\tilde\alpha_2^4\over 4}=0\;,
\end{equation}
where the interaction correction $\tilde\alpha_2$ in this case is given by 
\begin{equation}\label{correctionalpha2}
\tilde\alpha_2^4\equiv \alpha^4 \left| 1+{\varepsilon_b-z_{c,1}\over \varepsilon_a-z_{c,1}}\right|^2   \;.
\end{equation}
For the condition given in Eq.(\ref{cond1}b)
\begin{equation}
\tilde\alpha_2^4 > \alpha^4  \;,
\end{equation}
the effect of the interaction between the resonance states effectively increases the interaction of the bare discrete state with the continuum so that the value of the EP  (EP curve 1) which divides the stable-  and the single-resonance phases lies further from the continuum threshold  than  the single discrete state case (dashed line at  $\varepsilon_b=\varepsilon_{c,1}$ in Fig.\ref{fig:phase}):
\begin{equation}\label{EcApprox2}
0> \varepsilon_{c,1}>  \tilde\varepsilon_{c,2}\equiv-3\left({\pi \tilde\alpha_2^2\over 4}\right)^{2/3} -\tilde\alpha_2^2   \;.
\end{equation}

\begin{figure}[!]
\begin{center}
\includegraphics[width=8.5cm]{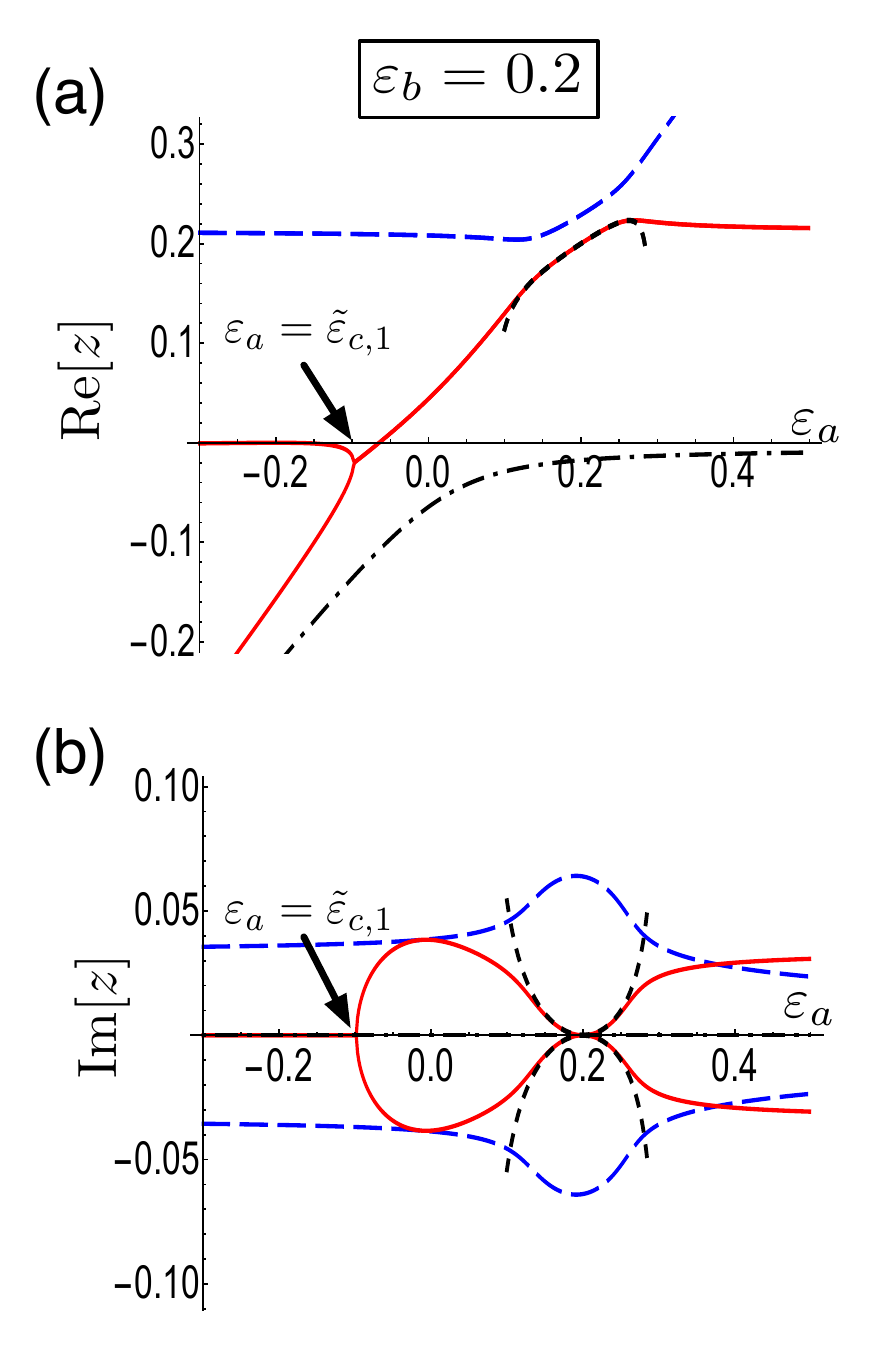}
\caption{(Color online) The eigenvalues of $\hat H_{\rm eff}$ as a function of $\varepsilon_a$ for a fixed value of $\varepsilon_b=0.2$.
Real parts and imaginary parts are shown in the upper and lower panels, respectively.
The resonant and anti-resonant states associated with  $|b\>$ are shown by the long-dashed curves, while the bifurcated solutions associated with $|a\>$ are shown by the solid curves.
The PBS is shown by the dashed-dotted curves.
The analytical expression around the Fano resonance given by Eq.(\ref{peAeS}) is represented by the short-dashed curves. 
The arrow indicates the EP. }
\label{fig:EV}
\end{center}
\end{figure}

In Fig.\ref{fig:EV},  we show the solutions of Eq.(\ref{Disp12}) as $\varepsilon_a$ varies with a fixed value of $\varepsilon_b=0.2$, where we have taken $\alpha=0.1$:  Real  and imaginary parts are shown in (a) and (b), respectively.
The change of $\varepsilon_a$ with the fixed value  $\varepsilon_b=0.2$ is indicated by a thin chained line in Fig.\ref{fig:phase}.
Since the dispersion equation (\ref{Disp12}) is a fifth-order polynomial, there are five solutions which are shown in Fig.\ref{fig:EV}.

On increasing $\varepsilon_a$ from far below the band edge, we encounter the EP  at $\varepsilon_a={\tilde \varepsilon}_{c,1}\equiv -0.099$.
 The behavior of the eigenvalues around the EP resembles that of the single discrete system shown in Fig.\ref{Fig:EVsingle}, because the discrete state $|b\>$ is energetically separated  from the $|a\>$ state so that the interaction between the two states is small.
The eigenvalues corresponding to the resonant and anti-resonant states associated with  $|b\>$ are shown by the long-dashed curves.
At this point the system transitions  from the single-resonance phase to the double-resonance phase, where the eigenvalues associated with the discrete state $|a\>$ bifurcate to form a resonance- and anti-resonance pair shown by the solid curves in Fig.\ref{fig:EV}.
The solution around the EP for the latter two states is approximately written as
\begin{equation}\label{Zpm}
z_\pm(\varepsilon_a)={\varepsilon_a+\tilde\alpha^2 \over 3}\pm {i\over \sqrt 3}(2\pi \tilde \alpha^2)^{1/3}\sqrt{\varepsilon_a-\tilde\varepsilon_{c}} \;,
\end{equation}
where $\tilde\alpha$ and $\tilde\varepsilon_{c}$ are $\tilde\alpha_1$ and $\tilde\varepsilon_{c,1}$ for the case of Eq.(\ref{cond1}a), and  $\tilde\alpha_2$ and $\tilde\varepsilon_{c,2}$ for the case of Eq.(\ref{cond1}b). 
The first derivative of the eigenvalues  is discontinuous at the EP  so that again the time-symmetry breaking happens  as a second order phase transition.
It should be emphasized that the  time-symmetry breaking is again non-analytically exaggerated by the Van Hove singularity as in Section~\ref{Sec:Single}.

As $\varepsilon_a$  further increases and comes close to $\varepsilon_b$, i.e. $\varepsilon_D\simeq 0$, the two decay channels from the $|a\>$ state and the $|b\>$ state interfere, resulting in the Fano resonance effect as mentioned in Section~\ref{Sec:Model}. 
In order to see the behavior of the eigenvalues more closely, we expand the solution around the Fano resonance:
\begin{equation}\label{z_approx}
z=\varepsilon_A+p(\varepsilon_A,\varepsilon_D)  \;,
\end{equation}
where $p(\varepsilon_A,\varepsilon_D)$ is a small deviation from $z=\varepsilon_A$ that vanishes as $\varepsilon_D=(\varepsilon_a-\varepsilon_b)/2\rightarrow 0$:  
\begin{equation}\label{plim}
 \lim_{\varepsilon_D\to 0}p(\varepsilon_A,\varepsilon_D)=0\;.
\end{equation}
Substituting Eq.(\ref{z_approx}) into Eq.(\ref{Disp_AS}) and neglecting   terms  higher than $p^2$, we obtain
\begin{eqnarray}\label{Disp_AS2nd}
&&\left(4\alpha^4 (\varepsilon_A+\varepsilon_D^2)-2\varepsilon_A \varepsilon_D^2+\pi^2\alpha^4\right)p^2  \nonumber\\
&&+\varepsilon_D^2(4\alpha^2 \varepsilon_A +\varepsilon_D^2) p+ \varepsilon_A \varepsilon_D^4=0 \;.
\end{eqnarray}
The solution of Eq.(\ref{Disp_AS2nd}) is given by
\begin{equation}\label{peAeS}
p(\varepsilon_A,\varepsilon_D)={-\varepsilon_D^2 (4\alpha^2 \varepsilon_A +\varepsilon_D^2)\pm \sqrt{ D(\varepsilon_A,\varepsilon_D)} \over 2 (4\alpha^4 (\varepsilon_A+\varepsilon_D^2)-2\varepsilon_A \varepsilon_D^2+\pi^2\alpha^4)}  \;,
\end{equation}
where
\begin{eqnarray}\label{EP_eAeS}
&&D(\varepsilon_A,\varepsilon_D) \nonumber\\
&=&-4 \pi^2\alpha^4 \varepsilon_A \varepsilon_D^4 \left( 1+ {2(\alpha^2-\varepsilon_A)\varepsilon_D^2 \over  \pi^2 \alpha^4} -{\varepsilon_D^4 \over 4 \pi^2 \alpha^4 \varepsilon_A} \right) \;. \nonumber\\
\end{eqnarray}
This solution well represents the exact solution around $\varepsilon_D\simeq 0$ as shown in Fig.\ref{fig:EV} by the short-dashed curves.
For  small $\varepsilon_D$, we expand Eq.(\ref{peAeS}) around $\varepsilon_D=0$ to yield
\begin{equation}\label{z_AS_expndS}
z=\varepsilon_A- {2 \varepsilon_A  \varepsilon_D^2 \over  \alpha^2 (\pi^2 + 4 \varepsilon_A)  } \pm i { \pi \sqrt{\varepsilon_A}\;  \varepsilon_D^2 \over  \alpha^2 (\pi^2 + 4 \varepsilon_A)  }  + O(\varepsilon_D^4) \;.
\end{equation}
The decay rate  quadratically  increases with $\varepsilon_D$ while it depends on $\sqrt{\varepsilon_A}$.
Therefore as $\varepsilon_A$ becomes small, the decay process is more suppressed and the state becomes quasi-stable in a wider parameter range. 

As shown in Fig.\ref{fig:phase}, as $\varepsilon_b$ approaches to the continuum threshold, the EP along the EP curve 2 shifts toward the band edge, and the EP and the Fano resonance  meet  at  $\varepsilon_a=\varepsilon_b=0$ ($\varepsilon_A =\varepsilon_D =0$).  
This is the point where  the Fano interference overwhelms the nonanalytical decay enhancement due to the Van Hove singularity, causing the TSBPT  to be drastically modified.
In order to see this, by taking the parameters as
\begin{equation}\label{eAeSTheta}
\varepsilon_D=\varepsilon \cos\theta\;,\;  \varepsilon_A=\varepsilon \sin\theta  \;,  \qquad (0\leq \theta \leq 2\pi)
\end{equation}
and substituting this  into Eq.(\ref{z_AS_expndS}), we find the eigenvalues expressed by 
\begin{eqnarray}\label{z_AStheta}
z&=& (\sin\theta) \varepsilon -{2\cos^2\theta \sin\theta \over \pi^2\alpha^2}\varepsilon^3\nonumber\\
&\pm&  i {i \cos^2\theta \sqrt{\sin\theta} \over\pi\alpha^2}\varepsilon^{5/2} +O(\varepsilon^{7/2}) \;,
\end{eqnarray}
under the condition 
\begin{equation}\label{epsilon1D}
\varepsilon\lesssim (4 \pi^2 \alpha^4)^{1/3}  \;.
\end{equation}
Therefore, as both  $\varepsilon_a$ and $\varepsilon_b$ approach the origin, the Fano resonance and the EP coincide as shown by the gray circle in Fig.\ref{fig:phase}, and the order of the phase transition becomes  fourth-order in the sense that the third-order derivative  for $\varepsilon$ is discontinuous.
We also note that the fractional power expansion, i.e. Puiseux expansion, of Eq.(\ref{z_AStheta})  starts with $\varepsilon^{5/2}$, different from the usual  behavior starting with $\varepsilon^{1/2}$ around the EP, which is revealed only by taking into account the nonlinearity of the eigenvalue problem of  the effective Hamiltonian.

\begin{figure}[t]
\begin{center}
\includegraphics[width=8.5cm]{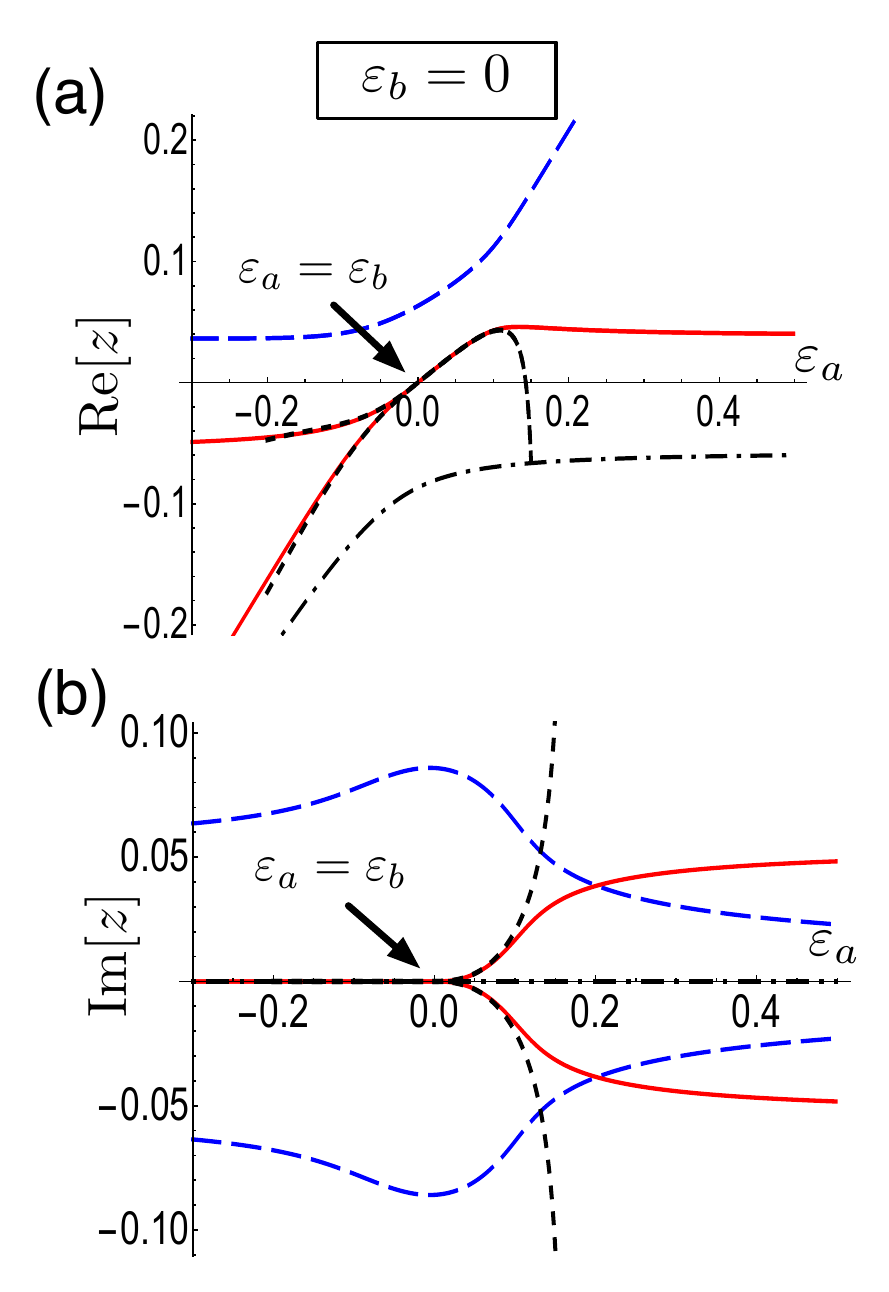}
\caption{(Color online) The eigenvalues of $\hat H_{\rm eff}$ as a function of $\varepsilon_a$ for a fixed value of $\varepsilon_b=0$.
Real parts and imaginary parts are shown in the upper and lower panels, respectively.
Each solutions is represented in the same style  as in Fig.\ref{fig:EV}.
The arrow indicates the EP. }
\label{fig:EPFano}
\end{center}
\end{figure}

We show in Fig.\ref{fig:EPFano} the exact solutions of Eq.(\ref{Disp12}) as $\varepsilon_a$ varies with a fixed value of $\varepsilon_b = 0$: Real and imaginary parts are shown in (a) and (b), respectively. 
The analytical approximation of the eigenvalues given by Eq.(\ref{z_AStheta}) is drawn by the short-dashed curve in Fig.\ref{fig:EPFano}, well reproducing the numerical results.
It is clearly seen that the order of the time-symmetry breaking transition becomes  fourth-order when the EP and Fano point meet together indicated by the arrows.

\begin{figure}[t]
\begin{center}
\includegraphics[width=8.5cm]{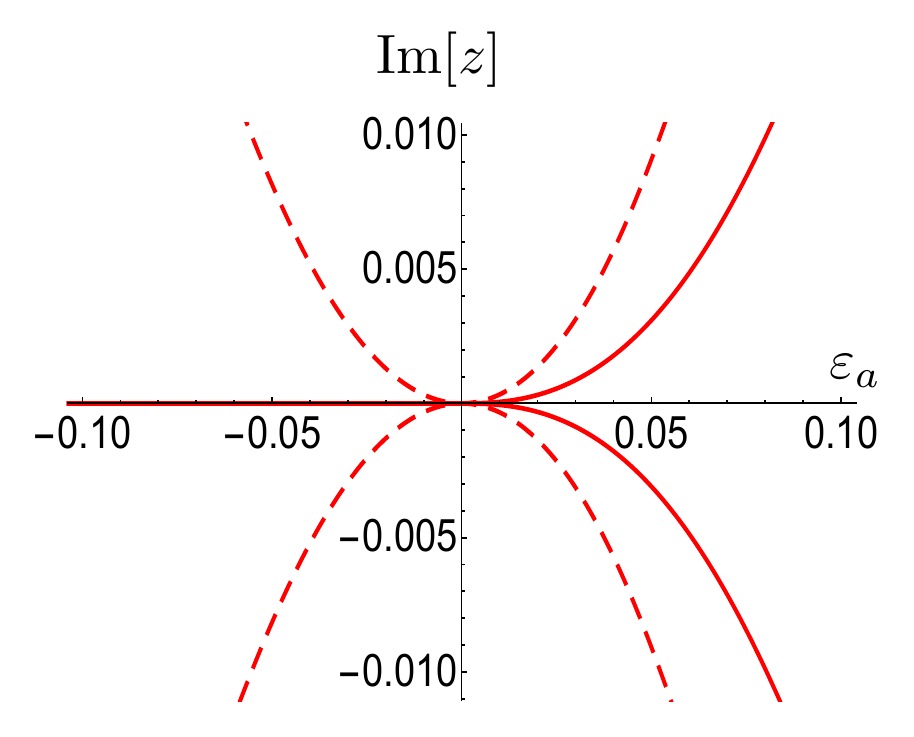}
\caption{(Color online) Comparison of the decay rates of Fig.\ref{fig:EV}(b) (dashed curve) and Fig.\ref{fig:EPFano}(b) (solid curve).
For the comparison, the decay rate of Fig.\ref{fig:EV} is shifted by $\varepsilon_b=0.2$  in the horizontal axis so that the Fano resonance of the both curves coincide at $\varepsilon_a=0$.
 }
\label{fig:decay}
\end{center}
\end{figure}

Furthermore, we find that the cooperation of the Fano resonance and the EP influenced by the Van Hove singularity makes the decaying state more stabilized than the ordinary Fano resonance.
In Fig.\ref{fig:decay}, we compare the decay rates of  Fig.\ref{fig:EV}(b) (dashed curve) and Fig.\ref{fig:EPFano}(b) (solid curve), where the decay rate of Fig.\ref{fig:EV} is shifted by $\varepsilon_b=0.2$  on the horizontal axis so that the Fano resonance of  both curves coincide at $\varepsilon_a=0$.
We find that the decaying state at the meeting point of the EP and the Fano resonance ($\varepsilon_b=0$: solid curve) is more stable (smaller decaywidth) than for the ordinary  Fano resonance ($\varepsilon_b=0.2$: dashed curve).
Since the eigenvalues around the Fano resonance is proportional to $\varepsilon_D^2$ as seen in Eq.(\ref{z_AS_expndS}), the second derivative of the eigenvalues in terms of $\varepsilon_D$ becomes a measure of the stability: The smaller the second derivative is, the more reduced the decaywidth  becomes.
Since the decay rate around the Fano resonance is represented by
\begin{equation}
\gamma(\varepsilon_D;\varepsilon_b)\equiv \left|{\rm Im}\; z \right|={1\over \pi\alpha^2} \sqrt{\varepsilon_D+\varepsilon_b} \; \varepsilon_D^2 \;,
\end{equation}
where we have used  $\varepsilon_D+\varepsilon_b \ll \pi^2/4$, the second derivative of $\gamma(\varepsilon_D;\varepsilon_b)$  at the Fano resonance ($\varepsilon_D=0$)  is given by
\begin{equation}
{\partial^2\over \partial^2 \varepsilon_D}\gamma (\varepsilon_D;\varepsilon_b)\Big|_{\varepsilon_D=0} = {2\over \pi \alpha^2} \sqrt{\varepsilon_b} \;.
\end{equation}
Therefore, we find that the decaying state at the meeting point of the EP and the Fano resonance ($\varepsilon_b=0$) is more stable than in the ordinary Fano resonance ($\varepsilon_b=0.2$), as shown in Fig.\ref{fig:decay}.

As in the single discrete state system studied in the previous section, we show in Appendix~\ref{AppSec:Jordan} that we can  introduce an effective non-Hermitian Hamiltonian which is represented by a Jordan block matrix at the EP.

\section{Discussion}\label{Sec:Discussion}

We have shown  that as a result of the competition between the effects of an EP and the Fano resonance, the TSBPT is modified as a higher-order transition in a system consisting of  two discrete states  coupled to a common 1D continuum.
The Van Hove singularity characteristic of 1D systems exaggerates this higher-order transition.
Here studying the TSBPT in a 3D system, we show that this higher-order phase transition of time-symmetry breaking is ubiquitous but the effect is not so prominent in the absence of the Van Hove singularity.

\begin{figure}[!]
\begin{center}
\includegraphics[width=8.5cm]{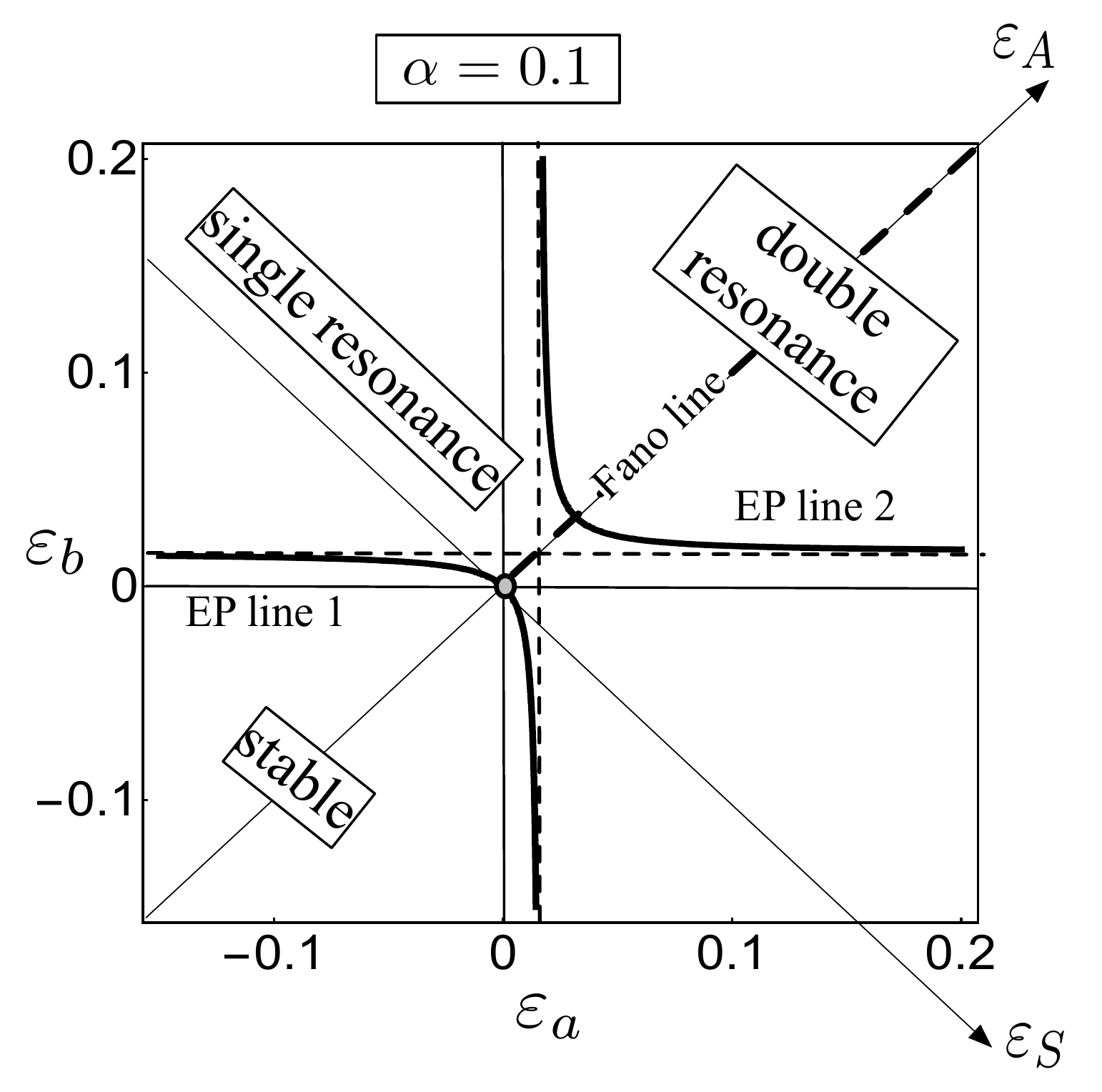}
\caption{Phase diagram for the time-symmetry breaking transition in a 3D system for $\alpha=0.1$.
Solid lines represent the EP as a function of $\varepsilon_a$ and $\varepsilon_b$.
A gray circle at the origin corresponds to the meeting point of the EP and the Fano resonance in Fig.\ref{fig:EPFano_3D}. 
Notations of the curves are the same as in Fig.\ref{fig:phase}.
}
\label{fig:phase3D}
\end{center}
\end{figure}

In a 3D system, the scalar-self energy in Eq.(\ref{selfEng})  is replaced by 
\begin{equation}\label{sigma3D}
\sigma_ (z)=-{\pi\over 2}\left(2+i \pi \sqrt{z}\right) \;,
\end{equation}
yielding the dispersion equation 
\begin{eqnarray}\label{Disp3D_AS}
f^{\rm{3D}}(z;\varepsilon_A,\varepsilon_D) &\equiv& \left\{ (z-\varepsilon_A) (z-\varepsilon_A+\pi \alpha^2)-\varepsilon_D^2 \right\}^2 \nonumber\\
&+&{\pi^4\alpha^4 \over 4}z   (z-\varepsilon_A )^2=0\;.
\end{eqnarray}
As in the preceding section, the EP curve is obtained by setting the resultant equal to zero, which is shown in Fig.\ref{fig:phase3D}.
It is found by comparison with Fig.\ref{fig:phase} for the 1D system that the EP curves (thick solid lines)  lies close to the EP curve of the single discrete state system (thin dashed line) in the 3D case.
This clearly shows that the effect of the interaction of the two discrete states is less pronounced in the 3D system than in the 1D system, because the Van Hove singularity enhances the interaction  in the 1D system.

In the 3D system, the higher order TSBPT occurs at the transition from the stable-phase to the single-resonance phase as denoted by the gray circle at $\varepsilon_a=\varepsilon_b=0$ ($\varepsilon_A=\varepsilon_D=0$).
As in the previous section, expanding $z$ around $\varepsilon_A$, i.e. $z=\varepsilon_A+p$, and leaving the terms up to second order in $p^2$ yields
\begin{equation}
(-2 \varepsilon_D^2 +  \pi^2  \alpha^4+ { \pi^4\alpha^4\over 4} \varepsilon_A) p^2 - 2\pi \alpha^2 \varepsilon_D^2   p +\varepsilon_D^4=0 \;.
\end{equation}
The solution is given by
\begin{equation}\label{p3D}
p={\pi \alpha^2 \varepsilon_D^2 \pm \sqrt{ -(\pi^4\alpha^4/4) \varepsilon_D^4 \varepsilon_A+2 \varepsilon_D^6 }
\over
-2 \varepsilon_D^2+\pi^2\alpha^4+(\pi^4\alpha^4/4)\varepsilon_A  } \;.
\end{equation}
Taking the variables given in Eq.(\ref{eAeSTheta}), Eq.(\ref{p3D}) reads
\begin{eqnarray}
p&=&{1\over \pi^2\alpha^4+{\pi^4\alpha^4\sin\theta \over 4} \varepsilon - 2 \cos^2\theta\varepsilon^2 } \nonumber\\
 &&\times \Bigg( \pi\alpha^2 \cos^2\theta \varepsilon^2  \nonumber\\
 && \pm i {\pi^2\alpha^2\cos^2\theta\sqrt{\sin\theta}\over 2} \varepsilon^{5/2}\sqrt{ 1-{8\cos^2\theta\varepsilon^2 \over \pi^4\alpha^4 \sin\theta } }  \Bigg) \;.
\end{eqnarray}
Under the condition 
\begin{equation}\label{Cond3D}
\varepsilon\lesssim {\pi^4 \alpha^4 \over 4}  \;,
\end{equation}
which is siginificantly limited in range compared to the 1D case as shown in Eq.(\ref{epsilon1D}), the solution of the dispersion equation is approximated by
\begin{equation}\label{zAS3D}
z=(\sin\theta)\varepsilon+{\cos^2\theta\over \pi\alpha^2}\varepsilon^2\pm  i{\cos^2\theta \sqrt{\sin\theta}\over 2\alpha^2}\varepsilon^{5/2} +O(\varepsilon^{3}) \;.
\end{equation}
Here we again  have the higher-order TSBPT as in the 1D system given by Eq.(\ref{z_AStheta}), but it should be emphasized that the parameter range of $\varepsilon$ to observe this effect is very narrow as shown in Eq.(\ref{Cond3D}).

\begin{figure}[t]
\begin{center}
\includegraphics[width=8.5cm]{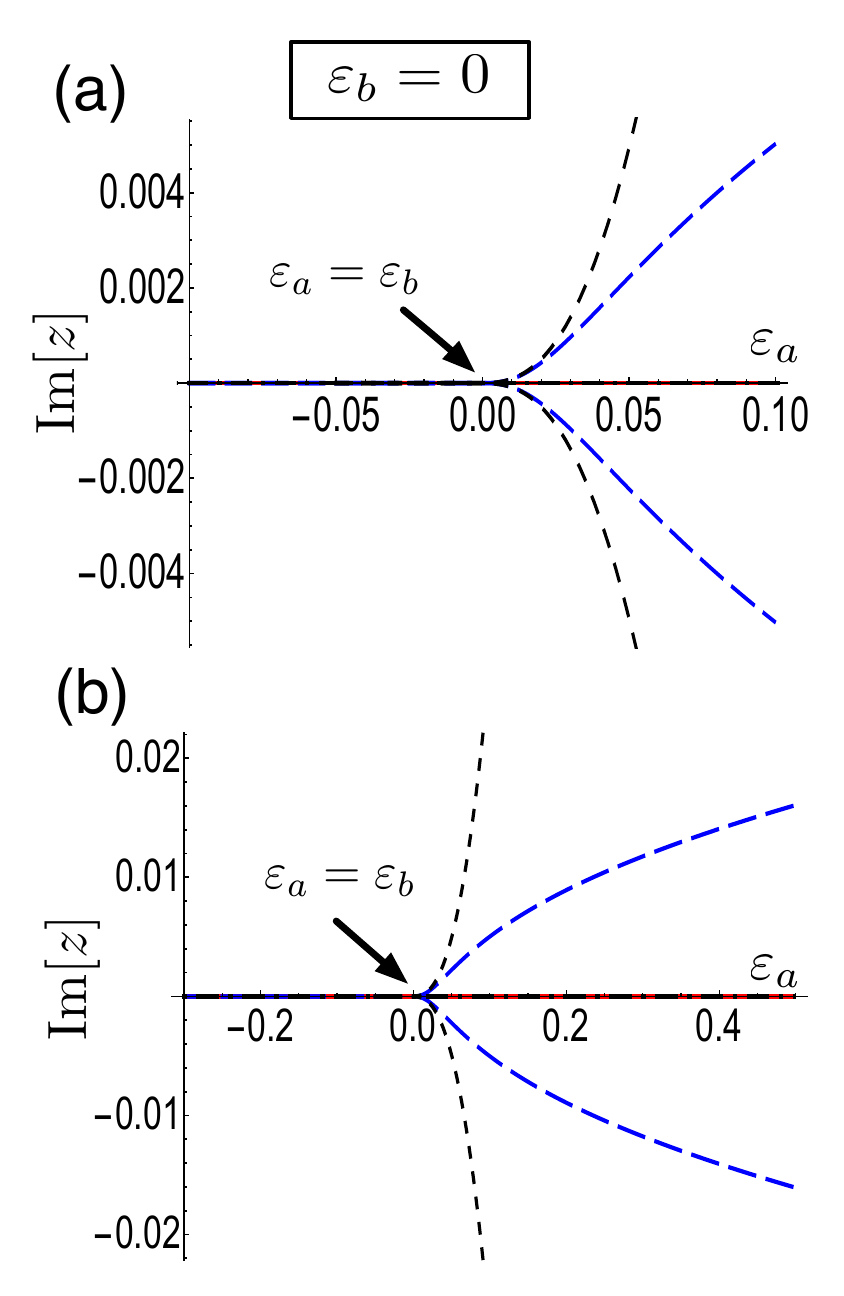}
\caption{(Color online) The imaginary part of the eigenvalues of $\hat H_{\rm eff}$ as a function of $\varepsilon_a$ for a fixed value of $\varepsilon_b=0$ in 3D system.
The bifurcated solutions associated with  $|b\>$ are shown by the long-dashed curves, while the analytical expression Eq.(\ref{zAS3D}) is shown by the short-dashed curves.
The arrow indicates the EP.
In (a), the fourth-order phase transition  is shown in a magnified scale around the EP, while in (b) the scale of the horizontal axis is the same as in Figs.\ref{fig:EV} and \ref{fig:EPFano}.
 }
\label{fig:EPFano_3D}
\end{center}
\end{figure}

In Fig.\ref{fig:EPFano_3D}, we show the imaginary part of the eigenvalues as a function of $\varepsilon_a$ for the fixed value  $\varepsilon_b=0$, where the EP and the Fano resonance coincide: we show (a)  a magnified scale, and (b) the same scale as in Fig.\ref{fig:EV}(b). 
Similar to the 1D system, the time-symmetry breaking transition occurs as a fourth-order phase transition.
However, the range of this smooth transition is very narrow compared to the 1D system, as seen in Fig.\ref{fig:EPFano_3D}(b).
This illustrates that the Van Hove singularity in the 1D system enhances the higher-order phase transition.

\begin{figure}[!]
\begin{center}
\includegraphics[width=8.5cm]{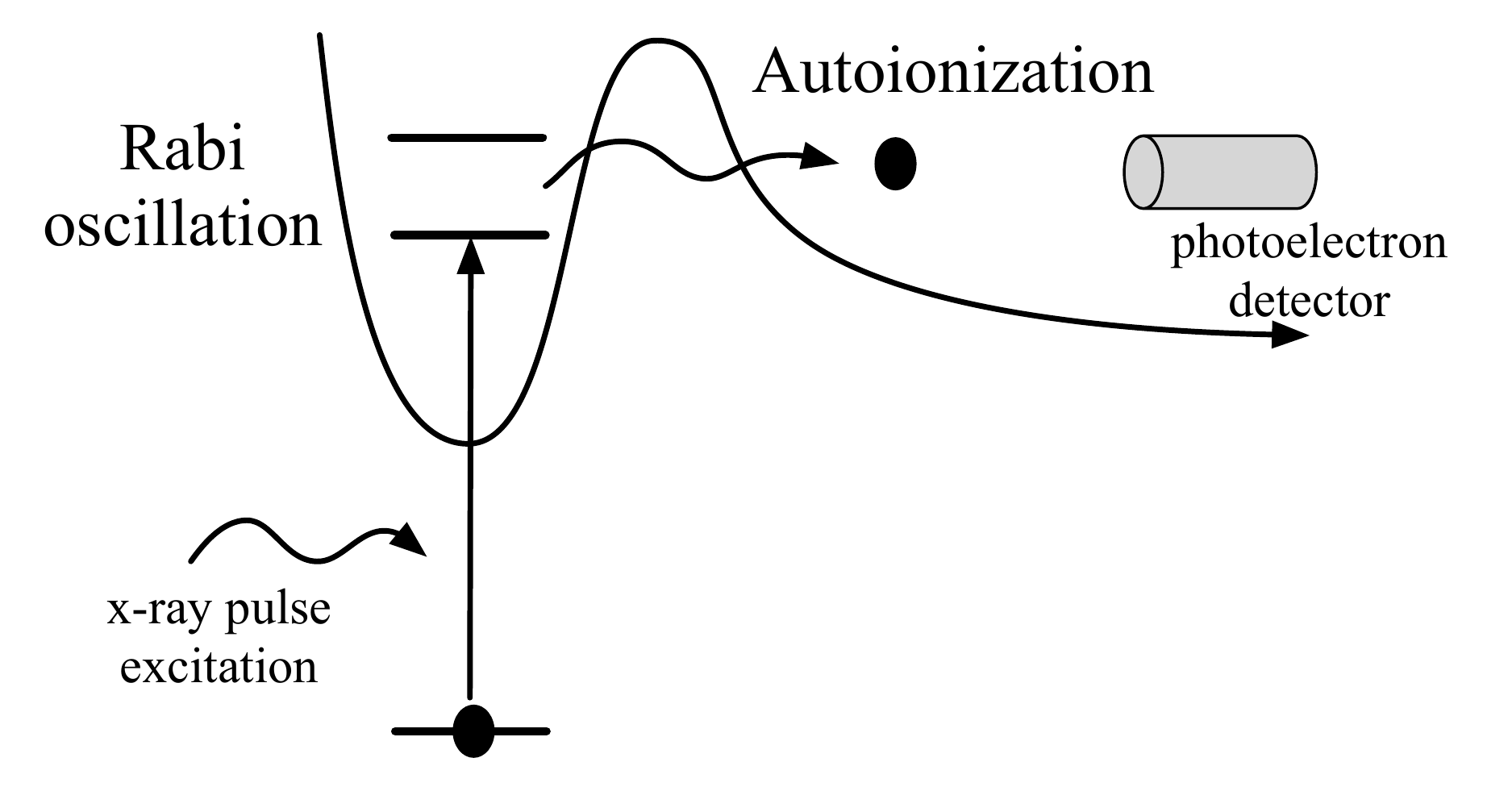}
\caption{Detection of a real time development of an autoionization decay of  an atom or a molecule with use of TRXAS and TRPES.}
\label{fig:TRXAS}
\end{center}
\end{figure}

The drastic change in the higher order TSBPT due to the cooperation of the EP and the Fano resonance at the continuum threshold shown in Fig.\ref{fig:EPFano} can be experimentally observed in the autoionization decay of an atom or a molecule with use of time-resolved ultrafast spectroscopy \cite{Shah, Femto,Tanaka01,Tanaka03}.
Here we  propose an experiment which uses a combination of  time-resolved x-ray absorption (TRXAS) \cite{Milne14} and  time-resolved photoelectron spectroscopies (TRPES) \cite{Suzuki06}, as shown in Fig.\ref{fig:TRXAS}, which should capture this characteristic phase transition.
In TRXAS,  an ultrashort x-ray pulse excites a core electron  to the discrete states near the ionized threshold, such as Rydberg states, and Rabi oscillation is induced between the resonance states by the pulsed excitation, which is observed by a delayed absorption probe.
The frequency of the Rabi oscillation corresponds to the difference in the real parts of the complex eigenvalues, and the damping rate of the Rabi oscillation reflects the decay rate due to the autoionization of the excited electron into the continuum.  
In TRPES, the autoionized photoelectron is detected by a time-resolved detector. 
Since  TRPES directly detects the decay product of the photoelectron, it reflects the imaginary part of the eigenvalues much more clearly than the damped Rabi oscillation by TRXAS. 
For example, the oscillatory behavior of  TRPES corresponding to the Rabi oscillation is drastically terminated at the Fano resonance  because one of the decay channels is completely suppressed.
Therefore when we measure both  TRXAS and TRPES and compare them, we can get a full picture of the decay process including the TSBPT.
Detailed  theoretical analysis of these spectroscopic experiments is now under study.

\acknowledgements

We thank K. Noba,  N. Hatano,  C. Uchiyama, K. Mizoguchi, and Y. Kayanuma for insightful discussions. 
The research of  S. G. was partially supported by a Young Researchers Grant from Osaka Prefecture University and the Program to Disseminate Tenure Tracking System, MEXT, Japan.

\appendix
\section{Scalar self-energy}\label{AppSec:FPO}

\begin{figure}[t]
\begin{center}
\includegraphics[width=8.5cm]{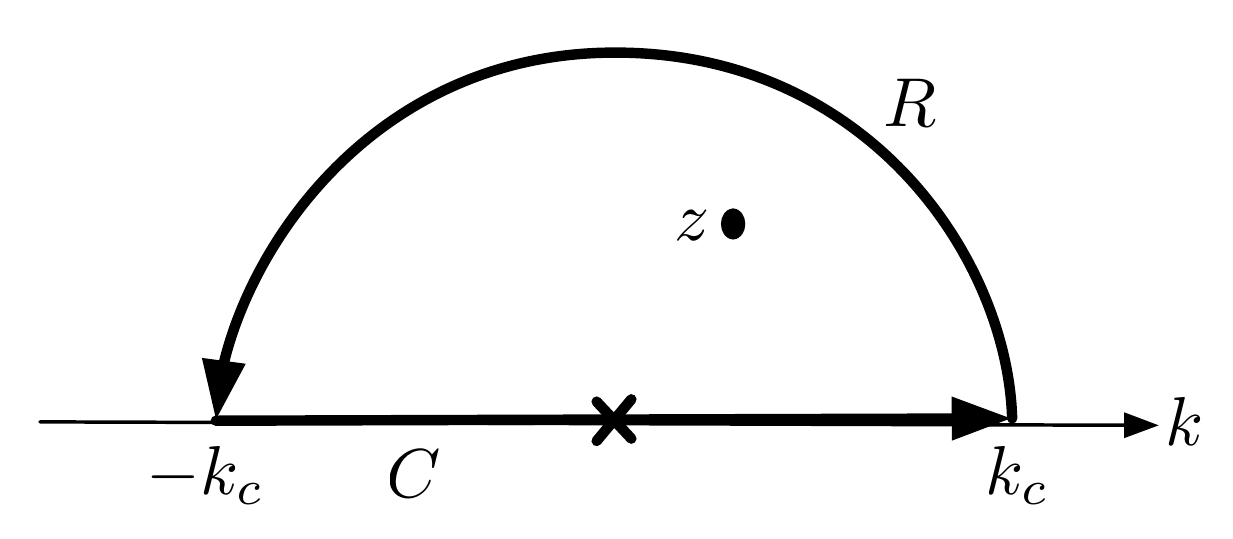}
\caption{Contour for the integral Eq.(\ref{sigmaCont}).   }
\label{Appfig:Int}
\end{center}
\end{figure}

The scalar self-energy for the 1D system is calculated by
\begin{equation}\label{App:sigma1D}
\sigma(z)={1 \over 2 k_c}\int_{-k_c}^{k_c}{dk \over z-k^2}  \;,
\end{equation}
where we assume
\begin{equation}\label{App:zkc}
|z|\ll k_c  \;.
\end{equation}
We rewrite this in terms of  the contour integral shown in Fig.\ref{Appfig:Int}, so that
\begin{equation}\label{sigmaCont}
\sigma(z)=\left\{ \oint_C-\int_R\right\}{dk\over z-k^2} \;,
\end{equation}
where $\oint_C$ and $\int_R$ denote a closed contour and a semicircle contour in Fig.\ref{Appfig:Int}, respectively.

Taking the residue at $\sqrt{z}$,
\begin{equation}\label{App:IntC}
\oint_C {dk\over z-k^2}=2\pi i {\rm Res} (k=\sqrt{z})=-i{\pi\over \sqrt{z}} \;.
\end{equation}
For the contour integral  $R$,  by taking $k=k_c \exp[i\varphi]$, we have
\begin{equation}\label{App:IntR}
\int_R {dk\over z-k^2}=i k_c \int_0^\pi {e^{i \varphi} \over z-k_c^2 e^{2 i \varphi}}\simeq -{2\over k_c} \;.
\end{equation}
By Eqs.(\ref{App:IntC}) and (\ref{App:IntR}),
\begin{equation}
\sigma(z)={1 \over 2 k_c}\left( {2\over k_c}-{i \pi\over \sqrt{z}} \right) \;,
\end{equation}
which gives Eq.(\ref{selfEng}).

In the 3D system, we define the scalar self energy as
\begin{equation}\label{App:sigma3D}
\sigma(z)={1\over (2 k_c)^3}\int_{-{\bf k}_c}^{{\bf k}_c}{d^3{\bf k} \over z-k^2} = {\pi\over k_c^3} \int_0^{k_c}{k^2 dk\over z-k^2}  \;.
\end{equation}
In this case, the contour integral for $C$ is given by
\begin{equation}\label{App:IntC3D}
\oint_C {k^2 dk\over z-k^2}=2\pi i {\rm Res} (k=\sqrt{z})=-i\pi \sqrt{z} \;.
\end{equation}
and for $R$ as
\begin{equation}\label{App:IntR3D}
\int_R {k^2 dk\over z-k^2}=i k_c \int_0^\pi {k_c^2 e^{i \varphi} \over z-k_c^2 e^{2 i \varphi}}\simeq 2 k_c \;.
\end{equation}
From Eqs.(\ref{App:IntC3D}) and (\ref{App:IntR3D}), we have
\begin{equation}
\sigma(z)={\pi \over 2 k_c^3}\left(-\pi i \sqrt{z}-2 k_c \right) \;,
\end{equation}
which gives Eq.(\ref{sigma3D}).

\section{Jordan block at the exceptional point}\label{AppSec:Jordan}

In this section, we introduce a heuristic  non-Hermitian effective Hamiltonian  as a $2\times 2$ matrix that is represented by a Jordan block at the EP. (We shall then call it {\it a $2\times 2$  effective  Hamiltonian}.)
We show elsewhere more formally that the bifurcation point determined above is an EP at which  not only the eigenvalues but also the eigenfunctions coalesce, and as a result, the Hamiltonian takes the form of a Jordan block in terms of eigenstate and pseudo-eigenbasis \cite{KatoBook,Hatano14}.

Let us write an effective two-by-two Hamiltonian given by 
\begin{equation}\label{App:BarH}
{\cal H}_{2\times 2}=
\begin{pmatrix}
a(\varepsilon_a) & 1 \\
-b^2 c(\varepsilon_a)  & a(\varepsilon_a)
\end{pmatrix}  \;,
\end{equation}
where the matrix elements are defined by
\begin{subequations}
\begin{eqnarray}
a(\varepsilon_a)&\equiv&{\varepsilon_a+\alpha^2 \over 3} \;,\\
b&\equiv&{(2\pi\alpha^2)^{1/3}\over \sqrt 3} \;,\\
c(\varepsilon_a)&\equiv&\varepsilon_a-\varepsilon_{c,1} \;.
\end{eqnarray}
\end{subequations}
Here we have taken ${\cal  H}_{2\times 2}$ such that the matrix elements are not singular in terms of the system parameter $\varepsilon_a$.

The eigenvalues are obtained as the solutions of the dispersion equation
\begin{equation}
z^2-2 a(\varepsilon_a) z + a^2(\varepsilon_a) + b^2 c(\varepsilon_a)=0 \;.
\end{equation}
yielding  the eigenvalues as
\begin{equation}
z_\pm=a(\varepsilon_a) \pm i b\sqrt{c(\varepsilon_a)} \;,
\end{equation}
which are the same as Eq.(\ref{SingleZpm}).
It is obvious from Eq.(\ref{App:BarH}) that  ${\cal  H}_{2\times 2}$ is undiagonalizable at the EP for $c(\varepsilon_a)=0$ because it takes a Jordan block structure.

For the two discrete system studied in Section \ref{Sec:Double}, we can again introduce a heuristic effective $2\times 2$ non-Hermitian  Hamiltonian given by
\begin{equation}\label{BarH_eff}
{\cal H}_{2\times 2}=
\begin{pmatrix}
a(\varepsilon_A,\varepsilon_D) & 1 \\
-b^2(\varepsilon_A,\varepsilon_D) c(\varepsilon_A)  & a(\varepsilon_A,\varepsilon_D)
\end{pmatrix}  \;,
\end{equation}
where we denote
\begin{subequations}
\begin{eqnarray}
a(\varepsilon_A,\varepsilon_D)&=& \varepsilon_A- {2 \varepsilon_A  \varepsilon_D^2 \over  \alpha^2 (\pi^2 + 4 \varepsilon_A)  }  \;,\\
b(\varepsilon_A,\varepsilon_D)&=& {\pi  \varepsilon_D^2 \over  \alpha^2 (\pi^2 + 4 \varepsilon_A)  } \;,\\
c(\varepsilon_A)&=&\varepsilon_A \;.
\end{eqnarray}
\end{subequations}
which gives the same eigenvalues as  Eq.(\ref{z_AS_expndS}).
It is obvious that at the EP  ($\varepsilon_A=0$)  ${\cal  H}_{\rm eff}$ is represented by a Jordan block matrix whose eigenstates coalesce as well as the eigenvalues.

While we have heuristically obtained an effective  Hamiltonian which takes the Jordan block form at the EP here, it can be derived from the microscopic dynamics by properly taking into account the component of the continuum subspace represented in Eq.(\ref{Qcomp}).
Strictly speaking, it is difficult to construct the Jordan block matrix at the EP just from knowledge of the effective Hamiltonian in the  subsystem represented by $\hat P$ in Eq.(\ref{Projection_ab}).
This can be done only when we deal the eigenvalue problem of the effective Hamiltonian consistent with that of the total Hamiltonian.
Since a thorough study of the eigenstate at the EP is beyond the scope of the present paper, it will be discussed in the forthcoming works \cite{Kanki16,Garmon16}.


\begin{thebibliography}{}
\bibitem{EP2A}
This point is an EP2A in the language of Ref.\cite{Garmon15}
\end{thebibliography}
\end{document}